\documentclass{article}

\PassOptionsToPackage{sort&compress}{natbib}
\usepackage{PRIMEarxiv}
\usepackage{natbib}

\usepackage[utf8]{inputenc} 
\usepackage[T1]{fontenc}    

\usepackage{url}            
\usepackage{booktabs}       
\usepackage{amsfonts}       
\usepackage{nicefrac}       
\usepackage{microtype}      
\usepackage{lipsum}
\usepackage{fancyhdr}       
\usepackage{graphicx}       
\graphicspath{{media/}}     

\usepackage{amsmath}
\usepackage{amsthm}      
\usepackage[table]{xcolor} 
\usepackage{bbm}
\usepackage{multirow}
\usepackage{longtable} 
\usepackage{caption} 

\usepackage[colorlinks=true, citecolor=blue, linkcolor=black, urlcolor=blue]{hyperref}

\title{RSDM: The Consensus Honest Money in the AI Era}

\author{
  Boliang Lin\textsuperscript{1,2} \\
  \textsuperscript{1}School of Traffic and Transportation \\
  Beijing Jiaotong University  \\
  Beijing 100044, China \\[2pt]
  \textsuperscript{2}Beijing Laboratory of National Economic \\
  Security Early-warning Engineering \\
  Beijing 100044, China \\[2pt]
  \texttt{bllin@bjtu.edu.cn} \\
  \And
  Ruixi Lin\textsuperscript{3} \\
  \textsuperscript{3}Independent Researcher\\
  Beijing 100044, China \\
}

\begin{document}
\maketitle

\begin{abstract}

 The medium of exchange of the traditional economy is mainly the fiat currency of each country or region, and when cross-border transactions occur, they need to be settled according to the exchange rate. In the AI world, however, the medium of exchange tends to be a globally recognized currency. Especially when AI acts as an agent for cross-border capital pool and cross cyclical asset allocation, it needs a sound money that can resist the depreciation of fiat currency and store long-term value. Therefore, we propose a globally consensus and universally accepted monetary rule framework for the AI era. The devaluation of money runs through almost the whole process of history, from the weight reduction and purity decrease of metallic coin to the unanchored over-issuance of paper currency. Whether it is the periodic compulsory recoinage in medieval Europe or Gesell's stamp scrip, both are essentially mechanisms for taxing money holdings.  Unlike Gesell's stamp scrip, \textbf{\textit{R}}edeemable \textbf{\textit{S}}elf-\textbf{\textit{D}}ecaying/Devaluing \textbf{\textit{M}}oney (RSDM) is a tokenized commodity money.Its essential innovation is to fill the hole in the storage fee of metal coins through the self-devaluing of metal weight recorded on the deposit certificate (warehouse receipt) of metal coins. In a sense, RSDM is an innovative version of \textit{Jiaozi} (a deposit receipt for base metal coin that emerged in Sichuan, China, about a thousand years ago). In this paper, we propose five forms of online and offline issuance of RSDM, providing a prototype for creating a globally recognized modern honest money.\\


\end{abstract}


\section{Introduction}
With the rise of the AI agent economy, more and more transactions will shift from ``human-operated payment'' to ``AI agent payment''. The transaction medium in the AI world needs to fulfill the functions of high-frequency, micro-amount, programmable, and globalized transactions for machines.

\indent An AI agent may call services in Country A today, purchase data from Country B tomorrow, and utilize computing power from Country C the day after. The medium of exchange inherently demands globalization. Traditional payments in the form of fiat currency (digital fiat currency or fiat-collateralized stablecoin) are inherently regionalized. Although stablecoins collateralized by US dollars, such as USDC, have become the default settlement currency for transactions in the AI world, the machine economy will, in the long run, require a globally recognized currency untethered from the sovereign credit of any single nation. Especially when people need to allocate assets across borders and cycles, they require an honest money that can hedge against fiat currency devaluation and serve as a long-term store of value, acting as the ultimate settlement currency for cross-chain and cross-platform transactions. By contrast, stablecoins collateralized by fiat currencies like the U.S. dollar are essentially digital proxies and shadow currencies of traditional fiat on the blockchain.

Currently, the rapid development of cryptocurrencies has exhibited characteristics of a supranational currency. BTC, in particular, is regarded as the ``digital gold'' \citep{taskinsoy2021bitcoin} of the AI Era, with some holders using it as a digital reserve asset and the ultimate value anchor against the over-issuance of fiat currency. There are currently over ten thousand cryptocurrencies in the global market \citep{Statista2024}. Given that cryptocurrencies lack intrinsic value support, they cannot serve as a stable measure of value in the long run and are incapable of shouldering the role of the ultimate universal currency of the digital age.

We believe that the medium of exchange in the AI Era needs to possess globally recognized intrinsic value, with long-term stable purchasing power and low circulation costs. Clearly, precious metals possess intrinsic value and are classic honest money. However, the major drawbacks of metallic weight-based money are high logistics costs, inconvenient storage and carrying, and for small-value transactions, they are difficult to accurately divide by weight. Especially for large-value transactions, carrying metal coins over long distances for trading poses significant security risks and incurs high transportation costs. Therefore, how to symbolize metal-weighed currency has always been a pain point in the history of currency evolution. For example, the ``Jiaozi'' \citep{battilossi2020handbook} that emerged in Sichuan, China, about a thousand years ago, the silver notes issued by Shanxi draft banks in China a few hundred years ago \citep{wu2018did}, the freely redeemable banknotes, and the receipts issued by early European goldsmiths to depositors are all examples of such efforts. However, all of them failed without exception. There are many apparent reasons, but the fundamental one is that the storage cost of metal cannot be dynamically recovered \citep{lin2025}. The RSDM designed by \citet{lin2016,lin2021,lin2025} meets the requirements of such a ultimate universal money. And the current programmable currency technology makes it easy to achieve the self-decay of RSDM.

The essence of currency is the claim to future value that has gained social consensus. Tokens that can be redeemed for metal coins are typical representatives of such claims \citep{gatch1996redeem}. For example, PAXG issued by Paxos Trust, XAUT issued by Tether, KAG issued by Kinesis, and so on \citep{zhang2025sok,harvey2025tokenized,oudang2026legal}, as well as the historically existing ``Jiaozi'' and freely redeemable banknotes. They only differ in terms of whether they are online or offline. However, the fatal flaw of these tokens (value claim vouchers) is the absence of a self-decay mechanism based on weight. Although issuers charge a fixed processing fee in addition to a fee based on the weight ratio of the metal when redeeming metal coins, these fees are unrelated to the length of time the collateral is stored. According to the mathematical proof in reference \citep{lin2021}, theoretically, as long as time is sufficient, such tokens will inevitably end up as credit-based currency or result in the bankruptcy of the issuing institution.

In fact, during the traditional economic age, people have always aspired for a global currency. However, with the advent of the AI Era, cross-border transactions have become more frequent, making the pursuit of a global currency even more urgent. The advantage of a global currency is that it can eliminate exchange rate risks and reduce transaction costs. A global currency must possess intrinsic value underpinned by worldwide consensus. For example, the Spanish silver dollar from the 17th to 19th centuries, the Bank of England notes that were freely convertible into gold, and the US dollar during the Bretton Woods system, all once approached a global consensus as currencies \citep{irigoin2020rise}.

As an honest, globally accepted symbolic money (either digital or paper-based), it should be easily redeemable for collateral (anchor assets) worldwide. Furthermore, the intrinsic value of such collateral should be historically tested and globally recognized. Additionally, the logistics cost of the collateral should be low. For example, the logistics costs of commodities such as electricity, iron ore, and crude oil are much higher than those of precious metals, making them unsuitable as good anchors. If electricity tokens are used as a global currency, due to the difficulty of cross-border power transmission, it would be challenging for electricity tokens issued by Country A to be redeemed and used in Country B. Therefore, the globalization of electricity tokens faces considerable difficulties, unless the emergence of high-energy density batteries in the future makes low-cost transportation of energy packets possible.

To sum up, in the AI Era, the globally accepted currency must be an honest money whose unit of value measurement can reflect the intrinsic value of money, such as weighed metallic currency and electricity currency measured in kilowatt-hours (kWh). Furthermore, such symbolic currency shall be redeemable for its agreed anchor assets, namely, it must feature full redeemability. Meanwhile, the logistics cost of this collateral should be low. For instance, the equivalent of a middle-class household's annual income denominated in such anchor assets can be easily stored at home. Therefore, the design of a sound money for the AI era remains a core and underexplored research topic in monetary theory.

Our key messages are as follows:
\begin{itemize}
    \item[(1)] A rule framework for globally consensual and universally acceptable currency in the AI era has been proposed.
    \item[(2)] Five online and offline issuance models for RSDM have been designed, providing a prototype for building a modern honest money with global consensus.
    \item[(3)] We also explored the technical approaches for banks to accept deposits and grant loans in RSDM.	
\end{itemize}
The remainder of this paper is organized as follows. Section \ref{sec:features} analyzes the core characteristics of medium of exchange in the AI era. we analyzes the historical depreciation process of monetary value in Section \ref{sec:history}. Section \ref{sec:compare} compares the similarities and differences between RSDM and Gesell's stamp scrip across fourteen aspects. In Section \ref{sec:framework}, we propose the framework of rules for globally consensus-accepted currency in the AI Era. Section \ref{sec:discuss} discusses five issuance models of RSDM and the channels for the banking system to absorb and grant RSDM loans. The final section presents the research conclusions.

\section{Monetary features and forms in the AI Era}
\label{sec:features}
At present, the mainstream strategy for trading in the AI world is to separate savings and payments: using Bitcoin for long-term value storage and stablecoins for daily payments. The motivation for this division is to avoid the impact of virtual currency price fluctuations on daily transactions, while also ensuring the long-term safety of assets. The result is that the form of currency has shifted from a single sovereign fiat currency to a composite system with multiple forms of currency coexisting.

\subsection{Key features of payment in the AI world}
We highlight key features of payment in the AI world as:
\begin{itemize}
    \item[(1)] Machine to machine (M2M) payments have become mainstream: AI agents settle autonomously between themselves, no longer relying on human operation or authorization.
    \item[(2)] Micro and high-frequency payments have become normalized. Billing is charged on a per-use, traffic-based, computing-power-based, or token-based basis. The single transaction amount is extremely small, the number of transactions is extremely large, and the settlement is in milliseconds.
    \item[(3)] Programmable payments (smart contract payments), automatic account splitting, automatic performance, and more.
\end{itemize}

\subsection{Consensus global currencies in the AI Era }
The essential feature of currency in the AI era is symbolization. Currencies shall be anchored to historically validated commodities with globally recognized intrinsic value. Full redeemability must be guaranteed. Currency itself has intelligent and programmable functions, which can exist in the form of code on the blockchain or in the form of paper currency for offline payments, ensuring that transactions can be carried out both online and offline scenarios.

\subsection{Analysis of major settlement currencies in the AI world}
At present, the settlement currencies in the AI world mainly include the following:
\begin{itemize}
    \item[(1)] \textbf{Fiat-Collateralized Stablecoins.} Stablecoins, due to their compliance and liquidity advantages, are suitable for high-frequency and cross-border transaction needs in AI scenarios. However, their main drawback is that stablecoins rely on centralized institutions for issuance and supervision. 
    \item[(2)] \textbf{Central Bank Digital Currencies (CBDCs).} As a medium of exchange for the AI world, the main advantages of CBDC are sovereign credit, programmability and supervisability. Its main disadvantages include centralized control, weak privacy and anonymity, inconsistent standards and functions of national CBDCs worldwide, and the need for currency exchange and settlement for AI cross-border transactions, which cannot adapt well to the AI-native borderless economy. It is difficult to form a globally unified universal monetary system in the short term.
    \item[(3)] \textbf{Commodity-backed stablecoins.} The advantages of commodity-backed stablecoins lie in low logistics costs and global historical consensus, making them an ``honest money'' in the AI world. For example, stablecoins issued 1:1 with precious metals as collateral resolve the drawbacks of physical gold as a medium of exchange, including difficulty in division and high logistics costs. Precious metals have served as hard currency for thousands of years. They are independent of sovereign credit, inflation-resistant, well-suited for long-term contracts, and store value across generations. Their major drawbacks include high custody costs for collateral, price fluctuations when precious metals are denominated in fiat currencies. Redemption is not convenient and cannot be exchanged in seconds like fiat-backed stablecoins. Currently, several mainstream stablecoins collateralized with precious metals do not charge warehousing fees based on the length of time. According to the proof in reference \citep{lin2021}, this redeemable token that does not dynamically charge storage fees based on the length of time is economically unsustainable.
    \item[(4)] \textbf{Power token.} Power coin is a resource-based token pegged to electricity itself and serves as an ownership certificate of energy. It represents the right to use a fixed amount of electric power, with 1 Powercoin generally equivalent to 1 kilowatt-hour (kWh) of physical electricity. Power token is regarded as the underlying asset of the new gold standard in the AI era, Powercoin still has notable shortcomings. Electricity cannot be stored or transmitted over long distances on a large scale at low cost. In addition, incompatible grid standards across nations make direct cross-border payment or charging via Powercoin practically unfeasible.
    \item[(5)] \textbf{Computational power token.} Computing power tokens are pegged to real computing resources, with their core mechanism lying in the tokenization of computing power. The main disadvantage is the rapid iteration of hardware, which causes continuous depreciation of computational power.
    \item[(6)] \textbf{Cross-AI-ecosystem universal value reserve currency.} A typical representative is Bitcoin. Its advantages include decentralization, no centralized issuing authority, and anti-censorship, making it suitable for long-term holdings by AI agents and cross-system value transfer. In terms of drawbacks, it suffers from extreme price volatility. As a payment medium, it carries substantial exchange rate risks and cannot serve as a stable unit of account for long-term economic planning.
\end{itemize}

Based on the above analysis, the monetary system in the AI Era is an ecosystem featuring diversified coexistence and hierarchical complementarity. CBDC is a compliance bridge that connects AI with traditional finance. From a long-term historical perspective, only precious metals have reached a consensus on value. However, if precious metal-backed tokens cannot be democratized and redeemable, they will inevitably end up with the same fate as fiat credit currencies. Electricity-backed tokens can be regarded as a type of commodity money with excessively high logistics costs, such as those incurred by pumped storage power stations. Stablecoins based on precious metals are currently the most globally recognized and historically tested currency among several existing payment mediums, but they require overcoming the black hole problem of dynamic storage costs.

\section{The historical depreciation process of monetary value}
\label{sec:history}
The history of money is a chronicle of continuous depreciation in monetary purchasing power and relentless dilution of credit. From the reduction of weight and purity of metal coin to the process of "anchor free over issuance" of paper currency. From a logistics perspective, the evolution of currency is a steady decline in the logistics costs of money, including storage and transaction costs.In the process of monetary evolution, human beings always tend to choose money with lower logistics cost. Obviously, as a medium of exchange, the minimization of logistics cost is one of the goals pursued in the process of monetary development \citep{lin2021}.The emergence of representative money is a milestone to reduce logistics cost of commodity money.

\subsection{The debasement history of metallic currency}
The debasement of metallic currency (mainly gold, silver and copper coins) is mainly reflected in the fact that the nominal value of currency falls far below its intrinsic metallic value, resulting in a decline in purchasing power. Governments usually dilute the value of money in covert or overt ways to seigniorage and plunder private wealth, by means of reducing fineness (e.g., lowering precious metal content), cutting coin weight (issuing smaller or thinner coins while keeping the nominal denomination unchanged), overvaluing currency parities (mandating one new coin to be equivalent to multiple old coins), and excessive coinage (mass minting amid fiscal collapse, with the total currency supply far exceeding the demand for commodity circulation).

In the early Roman Empire, the silver Denarius \citep{pense1992decline,butcher2012beginning} contained 3.9 grams of pure silver with a fineness of 98\%. In 64 AD, during the reign of Nero, to fund the reconstruction of Rome, the silver content of the Denarius was reduced to 93\%, with pure silver weight dropping to 3.4 grams. This marked the pivotal beginning of moderate coin debasement in the Roman Empire. Entering the 2nd century AD, the successive reigns of Trajan and Hadrian continued to cut the silver content of the Denarius gradually from 93\% to 85\%. At the start of the 3rd century AD, Caracalla halved its silver content to 50\% and issued new coins, effectively expanding the overall money supply. By the reign of Gallienus in 260 AD, the silver purity plummeted to merely 5\%, reducing the coin to a copper-zinc alloy. In the late Roman Empire, the original Denarius completely lost its circulating value. Civil transactions regressed to barter trade or relied solely on gold coins.

Similar phenomena also occurred in ancient China. Take the \textbf{\textit{Five-zhu Coin}} \citep{qian2018defended,xiaoqiang2023foundry} minted during the reign of Emperor Wu of Han in 118 BC as an example, it contained 4 grams of copper, unified the national monetary system, and sustained monetary stability for over a century. During the Wang Mang regime (7–14 AD), the imperial government issued decrees to forcibly abolish the original Five-zhu coins and introduced overvalued high-denomination coins. These new coins were legally mandated to circulate at an exchange rate of one new coin for fifty Five-zhu coins. Such measures triggered severe currency debasement, forcing commercial transactions to regress to a barter economy \citep{chen2012good}.

In 1803, the French government promulgated the Franc Law \citep{kindleberger2015financial}, subsequently issuing gold and silver francs and mandating the abolition of old currencies, which were exchanged at official fixed rates, thereby causing the wealth of old currency holders to shrink.

After Mexico gained independence in 1821, the government issued the new Mexican peso, enforcing a mandatory 1:1 exchange ratio against Spanish silver coins. However, the silver content of the new currency was significantly reduced, enabling the government to plunder private silver in a disguised form \citep{pieper2015money,irigoin2020rise}.

The renovatio monetae \citep{naismith2019money}, a mandatory institution of periodic currency invalidation and recoining widely enforced in medieval Europe (c. 1000–1300 AD), is essentially an ancient mechanism for taxing monetary holdings.

Silvio Gesell (1862–1930) proposed a monetary theory that aimed to eradicate currency hoarding, accelerate monetary circulation, and cure economic crises by imposing a holding cost on money. The stamp scrip \citep{blanc1998free} he designed required banknotes to be periodically affixed with stamps in order to retain their legal tender status. In effect, Gesell's stamp scrip and medieval Europe's monetary recoining system were logically equivalent at the foundational level. They merely achieved the same goal through different mechanisms—recoining versus stamping. Renovatio Monetae was the medieval equivalent of the Gesell Tax. Gesell's stamp scrip was not created out of thin air, but rather revived and theoretically systematized an ancient monetary management wisdom. Renovatio Monetae was the medieval ancestor to Gesell's stamp scrip. Both sought to disrupt the store of value function of currency and compel its circulation by artificially creating a cost of holding. The only distinction lies in their mechanism and intent: medieval recoining was a one-time, fiscally oriented tax, whereas Gesell's approach was an economically targeted, gradual regulatory tool. Gesell's stamp scrip is essentially analogous to historical cases where governments forced the exchange of old money for new at official exchange rates. It is fundamentally no different from the overissuance of paper currency—all represent distinct forms of currency depreciation.

\subsection{The history of hyperdepreciation in paper currency}
About a millennium ago, China's Sichuan region was forced to use iron money because of the depletion of copper mines, a serious shortage of raw materials for casting copper money, coupled with the central government's monetary controls over border regions. Low in value yet heavy in weight, iron coins were extremely cumbersome to carry and severely hindered commercial activities. This predicament prompted Sichuan merchants to invent the Jiaozi to resolve the inconvenience of transporting iron coins.

The early privately issued Jiaozi was essentially a monetized warehouse receipt backed by 100\% reserve requirements, fully redeemable for iron coins. Later, the imperial authorities issued Jiaozi under a fractional reserve system, making the Northern Song Jiaozi the world's first official fiat paper money. Gradually, it degenerated from a sound credit currency into a tool for fiscal overspending. It ultimately suffered from hyperinflation and total monetary collapse. Local commerce reverted to iron coins and barter transactions, rendering the Jiaozi a classic historical case of excessive paper currency issuance in ancient times.

Cases of vicious currency devaluation are also common in modern history. For example, in Weimar Germany (1921-1923), hyperinflation, money as wallpaper, and the savings of the middle class were wiped out \citep{xiao2019money}. During the Republic of China, the drastic devaluation of fabi and gold yuan certificates (1937–1949) forced the public to revert to barter trade or privately circulate silver dollars \citep{campbell1954}. Hungary's currency devaluation from 1945 to 1946 triggered the most extreme hyperinflation in history, ultimately resulting in a full monetary reset \citep{siklos1989end}. In the early 2000s, the drastic devaluation of the Zimbabwean dollar \citep{hanke2009measurement} eventually forced the country to abandon its domestic currency and adopt a multi-currency system, in which the US dollar, South African rand, Chinese Renminbi and other foreign currencies circulated side by side. Severe hyperinflation in Venezuela \citep{barredo2024credit} from 2016 to 2021 rendered the bolívar virtually worthless.

Numerous similar examples exist and will not be listed one by one. The rapid depreciation of a nation's fiat currency invariably prompts the public to abandon domestic money and hoard gold, foreign currencies and physical commodities as safe-haven assets.

In conclusion, it is clear that weighed precious metal money such as gold and silver ingots have rarely experienced the drastic devaluation seen in paper money throughout history. It seems that precious metals are a historically tested value with global consensus.

\section{A comparative study between RSDM and Gesell's stamp scrip}
\label{sec:compare}
\citet{lin2025} proposes a \textbf{\textit{R}}edeemable \textbf{\textit{S}}elf-\textbf{\textit{D}}ecaying/devaluing \textbf{\textit{M}}oney (RSDM) and regards it as a modern honest currency. In this paper, we define weight-based money represented by gold, silver and copper as the traditional honest currency.

\subsection{Attempts at monetaryization of warehouse receipts and their logistics cost black hole}
Due to the high logistics cost of metal money (especially base metals), in order to facilitate large transactions, people have also tried to monetize various warehouse receipts to solve this dilemma. Typical examples include the Jiaozi that emerged in Sichuan Province, China, roughly a thousand years ago, the bank notes issued by Shanxi draft banks in China centuries ago, and the early receipts issued by European goldsmiths to depositors. In fact, the U.S. dollar during the Bretton Woods era and the early Special Drawing Rights (SDRs) shared similar functions. The common feature of these tokens is that the issuer promises that the noteholder can be converted into precious (base) metals at any time (or under certain conditions). That is to say, the issuer promises that these paper tokens can be converted into agreed amounts of metal moey such as gold or silver according to the face value of the tokens.

Why did all commodity money tokens such as Jiaozi and silver bank notes ultimately break their promise of redeemability for collateral assets at any time? From the perspective of the storage cost of metallic money, References \citep{lin2021,lin2025} analyze the logistics cost black hole arising from the symbolization of commodity money. The research finds that if metallic money is stored for a sufficiently long period, the accumulated custody fees for collateral will eventually exceed static settlement fees.

In other words, if the redemption fee for gold and silver is set statically---that is, fixed as a stipulated percentage of the face value of paper tokens (for instance, holders of Jiaozi were required to pay a 3\% custody fee when redeeming metallic money)---the commitment to fully redeemable paper currency is inherently illusory.

To cover the ever-growing deficit of storage costs, ancient metallic money custodian commonly lent out depositors' physical gold and silver, or even issued uncovered bank notes, so as to gain extra income beyond settlement fees. Part of such revenue was used to offset storage expenditures, and a portion was further paid to depositors as interest. Consequently, the full-reserve principle derived from warehouse receipt mechanisms degenerated into an implicit fractional reserve system, and honest money gradually evolved into credit money. Ultimately, all paper tokens of convertible commodity money, which combined value storage function with low circulation costs, were eliminated by history, unable to withstand the erosion of the logistics cost black hole.

\subsection{RSDM: Modern honest money with both value storage and low transaction costs}
A fully redeemable note must have the consideration of a storage fee that increases over time, like a warehouse receipt. However, if warehouse receipts, most of which are under real-name registration, are directly used as a medium of exchange, due to the uncertainty of handling fees when exchanging physical metal coins, warehouse receipts cannot be widely circulated as currency.

As a global currency in the AI era, it shall possess two essential attributes: the long-cycle value storage capacity of traditional honest money, such as precious metals, and the near-zero logistics cost characteristic of mode currency rn paper or digital currency. History and reality demonstrate that Fiat or credit currency, metallic tokens, as well as contemporary stablecoins and virtual currencies, cannot achieve this dual advantage simultaneously.

The RSDM proposed by Lin and Lin (2025) is precisely such a modern honest money with value storage function and nearly zero logistics costs. Its mechanism is to transfer partial value by means of the self-depreciating weight of RSDM collateral, so as to cover the daily custody costs of the collateral. It is precisely this self-devaluing mechanism of metallic money weight that lays the cornerstone for a fully redeemable modern honest currency. In a sense, RSDM is a monetized warehouse receipt. However, it differs from ordinary warehouse receipts, which cannot circulate as currency. RSDM is also distinct from traditional representative money, such as the ancient Chinese silver bank notes issued one to two centuries ago. In history, private banks commonly invested depositors' entrusted silver in risky ventures. Under such circumstances, if the issuing institution suffered investment losses, note holders would be unable to redeem their collateral.

The collateral of RSDM must be fully supervised by third-party institutions and shall not be used for any risky investment. Superficially, this seems to be a waste of assets. Some hold the view that gold stored in underground vaults as a monetary anchor asset is a barbarous relic. And, it is meaningless to dig gold out of one hole in the ground (mines) only to place it into another hole underground (vaults). But they may have forgotten that money is essentially a social custom, a trust that can be exchanged for goods or redeemed for precious metals. Gold bars quietly reserved in underground vaults serve as one of the cornerstones for building such trust. Therefore, the so-called “barbarous relics” stored in vaults are the most solid foundation for maintaining public confidence in honest money.

\subsection{Silvio Gesell's Stamp Scrip}
The German economist Silvio Gesell (1862–1930) argued that money should conform to natural laws, possessing a limited service life and value depreciation characteristics just like other commodities. By introducing money holding costs, Gesell compelled currency holders to put money into circulation as quickly as possible.

\subsection{Analysis of the differences between RSDM and Gesell's stamp scrip}
Stamp scrip is essentially a replica of the historical practice of reducing the weight or precious metal content of metallic money, such as the periodic recoinage in medieval Europe. However, RSDM is fundamentally different in nature, and their major distinctions are outlined in Table \ref{tab:diff}.

\begingroup
\centering
\setlength{\tabcolsep}{1.5pt}
\begin{longtable}{@{}cccc@{}}
\caption{Differences Between RSDM and Stamp Scrip}
\label{tab:diff}\\
\toprule
\rowcolor[HTML]{DEEAF6} 
\textbf{No.} & \textbf{Item} & \textbf{RSDM} & \textbf{Stamp Scrip} \\ \midrule
1 & Goals and Motives & \begin{tabular}[c]{@{}c@{}}To symbolize metallic money, \\ thereby integrating the dual \\ advantages of value storage \\ and low transaction costs of \\ honest money. It also aims to \\ revitalize idle precious metal \\ assets held by the private sector.\end{tabular} & \begin{tabular}[c]{@{}c@{}}Accelerate currency circulation, \\ stimulate economic demand, \\ counteract economic crises, \\ and weaken the value \\ storage function of money.\end{tabular} \\ \midrule
\rowcolor[HTML]{ECF4FF} 
2 & Historical Origin & \begin{tabular}[c]{@{}c@{}}Jiaozi in ancient China \\ a thousand years ago;\\ the monetization of warehouse \\ receipts (metallic money \\ deposit certificates).\end{tabular} & \begin{tabular}[c]{@{}c@{}}Periodic Recoinage \\ in Medieval Europe\end{tabular} \\ \midrule
3 & Implementation Measures & \begin{tabular}[c]{@{}c@{}}Let the weight of metal money \\ in the form of representative \\ money self-decaying (Automatic \\ Weight Reduction) daily\end{tabular} & \begin{tabular}[c]{@{}c@{}}Gesell's stamp scrip requires \\ periodic currency stamp tax payment.\end{tabular} \\ \midrule
\rowcolor[HTML]{ECF4FF} 
4 & Theoretical Basis & \begin{tabular}[c]{@{}c@{}}Offset the collateral custody \\ fees incurred by professional \\ custodian institutions.\end{tabular} & \begin{tabular}[c]{@{}c@{}}Enable money to bear the \\ “time depreciation cost” like \\ ordinary commodities, narrow the \\ storage cost gap between currency \\ and goods, and levy a Demurrage.\end{tabular} \\ \midrule
5 & Theoretical Defects & None & \begin{tabular}[c]{@{}c@{}}This theory contains major flaws. \\ Individuals hold money as \\ well as goods, and the cost \\ of holding it has been borne \\ by themselves. Additional storage \\ fees are unreasonable. Does an \\ individual hold 50kg grain or \\ 5g gold without timely consumption \\ and have to pay stamp duty?\end{tabular} \\ \midrule
\rowcolor[HTML]{ECF4FF} 
6 & Types of money & \begin{tabular}[c]{@{}c@{}}The symbolized form of \\ commodity money / weighed money.\end{tabular} & Fiat currency \\ \midrule
7 & Effects on monetary functions & Retain all functions of traditional money & \begin{tabular}[c]{@{}c@{}}Damages the store-of-value \\ function of money\end{tabular} \\ \midrule
\rowcolor[HTML]{ECF4FF} 
8 & Is it redeemable? & Yes & No \\ \midrule
9 & Coerciveness & \begin{tabular}[c]{@{}c@{}}Non-coercive, it is a \\ spontaneous market behavior.\end{tabular} & \begin{tabular}[c]{@{}c@{}}Coercive acts imposed \\ by the government\end{tabular} \\ \midrule
\rowcolor[HTML]{ECF4FF} 
10 & Rule sequence & \begin{tabular}[c]{@{}c@{}}Announce the rules of the \\ decaying money first, then issue \\ and sell RSDM.\end{tabular} & \begin{tabular}[c]{@{}c@{}}Gesell's scheme essentially \\ imposed temporary stamp \\ duty rules on existing \\ currency, rather than issuing \\ a new form of money.\end{tabular} \\ \midrule
11 & Application & None & \begin{tabular}[c]{@{}c@{}}Small-scale experiments \\ were conducted in \\ towns or communities.\end{tabular} \\ \midrule
\rowcolor[HTML]{ECF4FF} 
12 & Rate of depreciation & Determined by warehousing costs & \begin{tabular}[c]{@{}c@{}}There is no theoretical \\ basis for setting the \\ stamp duty amount.\end{tabular} \\ \midrule
13 & Alternative pathway & \begin{tabular}[c]{@{}c@{}}Fully collateralized; no \\ viable alternative pathway.\end{tabular} & \begin{tabular}[c]{@{}c@{}}It can be replaced by \\ currency over-issuance.\end{tabular} \\
\rowcolor[HTML]{ECF4FF} 
14 & Theoretical enlightenment & \begin{tabular}[c]{@{}c@{}}Constructing Honest Monetary \\ System in the AI Era\end{tabular} & Negative interest rate policy \\ \bottomrule
\end{longtable}
\par
\endgroup

\subsection{Effects comparison: fiat overissuance and stamp scrip}
Assume that the Consumer Price Index (CPI) at time $t_1$ is 100. At this moment, an individual holds exactly 100 units of fiat currency, which can purchase 1 unit of a basket of goods at a certain ratio. Under Gesell's stamp scrip system, the individual pays a stamp tax of 5 monetary units at time $t_2$. Assuming the price of every commodity in the basket remains unchanged (a counterfactual premise), the remaining currency balance is only sufficient to purchase 0.95 units of the same goods basket.

By shifting perspective, an analogous outcome can also be achieved through 5.3\% excessive currency issuance. But the former is an explicit devaluation, and the latter is an implicit devaluation. In fact, as a credit currency, it is impossible to determine the logistics cost of its potential purchase of goods. Because of a unit of fiat currency, the number of baskets of goods purchased at different times cannot be determined. Moreover the inventory ratio and storage cost vary greatly across different commodities. For example, the cost of stockpiling one kilogram of gold is far lower than that of storing grain of equal value. It is impossible to determine how much of a unit of fiat currency goes to buy gold and how much to buy grain.

One viable solution is to quantify the overall inventory cost of commodities across the entire society. Numerous factors affect commodity inventory costs, including the depreciation of stocked goods, spoilage losses of food products, and the appreciation of certain commodities. In addition, merchants usually incorporate inventory costs into commodity selling prices. By contrast, state strategic reserves such as grain rely on taxation or currency devaluation to offset persistent inventory expenses. Accordingly, determining a reasonable currency depreciation rate purely based on commodity inventory costs remains highly complex. For a rough estimation, suppose total inventory costs account for 4.6\% of GDP, among which the capital occupancy cost of inventory makes up 2.5\%; then the corresponding currency depreciation rate can be stabilized at approximately 2.1\%. This mechanism can be practically realized through moderate excessive currency issuance.

\section{Framework of rules for globally consensus-accepted currency in the AI Era}
\label{sec:framework}
In the AI era, the globally consensus-accepted currency needs to meet the needs of offline and AI world. It needs to be machine-measurable, human-perceivable, and market-priceable. In a sense, monetary history is a process of value symbolization, evolving from commodity money to symbolic currency. The evolutionary trajectory follows a clear sequence: universal equivalent, weighed money, metallic coinage, paper money, and electronic money. Throughout history, money has consistently advanced along the pathways of creditization, symbolization and digitization.

\subsection{An ancient honest money: weight-based money}
Weight-based money, also known as weight money, refers to a monetary form with no standardized shape, fixed weight or uniform fineness. Its value must be verified through weighing and fineness judgment during transactions. As the value of weight-based money equals the intrinsic value of its raw material, and restricted by the reserves of monetary materials, it is extremely difficult to over-issue, making it a classical type of honest money with relatively stable value. In particular, weight-based money backed by precious metals enjoys global value recognition and unimpeded cross-regional circulation. Its major disadvantages are low circulation efficiency and high transaction costs, and it is difficult to expand synchronously with GDP due to resource constraints.

\subsection{Main defects of classical weight-based honest money}
Summary of main defects of classical weight-based honest money is:
\begin{itemize}
    \item[(1)] High costs incurred by weighing and fineness verification.
    \item[(2)] High logistics costs in storage and transportation, together with substantial security risks.
    \item[(3)] For small-sum transactions, precious metals cannot be accurately divided by weight.
    \item[(4)] Weight-based precious metal money is often hoarded, which can easily lead to insufficient market liquidity.
\end{itemize}

\subsection{Analysis of the advantages and disadvantages of quantity-based currency}
Quantity-based currency, also known as denomination-based currency, refers to standardized currency issued uniformly by the state. It requires no weighing or fineness appraisal in transactions. Since its value depends on the issuer's credit, there is a risk of over issuance depreciation. Its major advantages include high transaction efficiency and flexible money supply. By regulating the aggregate money stock, it can adapt to the demands of economic expansion. Represented by fiat currency, such currency is typically backed by national sovereignty. It requires currency exchange for cross-border use and thus lacks global acceptability.

Weight-based money and quantity-based currency are two core forms in the history of monetary development, marking the critical evolution from reliance on the intrinsic value of physical commodities to dependence on issuer credit. The two differ systematically in terms of value foundation, usage mode, historical stage, and circulation efficiency.

\subsection{Monetary units that reflect a currency's intrinsic value}
At present, Token has become the mainstream workload measurement unit in the AI world. However, there is no unified standard for defining the intrinsic value of Token, leading to a lack of equivalent recognition across different models. For instance, the energy consumption and actual value of one Token on Platform A may differ from those on Platform B. Mere Token counting fails to reflect real value. Consequently, Token cannot serve as a universal measure of value and is ineligible to function as a globally unified monetary unit.

Since the essence of AI lies in the conversion process from electricity to computing power and ultimately to intelligence. The kilowatt-hour (kWh) serves as the most fundamental quantitative unit within this chain. Therefore, it is reasonable for the AI world to adopt the electricity cost unit (kWh) as the underlying pricing foundation. As a power unit, the kWh possesses monetary attributes including measurability, divisibility, storability and universality. Its value is rooted in electricity supply capacity and cannot be diluted by excessive currency issuance. Nevertheless, it has prominent limitations. Electricity prices are affected by time periods, geographical regions, fuel costs, policy subsidies and other variables, making global unified pricing unfeasible. Furthermore, electric power cannot be stored on a large scale for a long time, and it suffers from substantial losses during long-distance transmission. Electricity prices vary across countries, and the barriers to cross-border power transmission, therefore, the kilowatt-hour is far from an optimal option as the universal measure of value in the AI world.

Fiat currency, including CBDC and fiat-backed stablecoins, has no intrinsic value and no upper limit on the total amount. It is inherently prone to over-issuance and long-term depreciation, with high costs for cross-border conversion. Bound by sovereign constraints and exposed to the risks of financial weaponization, fiat currency can hardly serve as a globally recognized measure of value. It functions merely as a transitional tool and cannot evolve into a universal global currency in the AI era.

\subsection{Framework of rules for globally honest money in the AI Era}
Traditional money is essentially a matter of trust and social convention, while the medium of exchange in the AI world must satisfy key characteristics: machine measurable, human perceptibility, and market price discoverability. Integrating the attributes of both traditional money and on-chain currency, we establish the rule framework for a globally currency in the AI Era as follows:
\begin{itemize}
    \item[(1)] It must be a symbolic currency, such as paper currency or digital currency.
    \item[(2)] The monetary unit shall reflect the intrinsic value of the money, such as weight units for metals, gram, ounce and so on, and kilowatt-hour (kWh) for electric energy.
    \item[(3)] The anchor assets of money should be easily standardized commodities, such as metals such as gold, silver and copper.
    \item[(4)] The intrinsic value of anchor assets shall be historically validated and universally recognized on a global scale, such as electricity, grain and gold.
    \item[(5)] The collateral assets behind the currency shall be 100\% redeemable, and the redemption points are widely distributed around the world, so it is convenient for holders to redeem the collateral.
    \item[(6)] The currency's anchored assets shall feature low logistics costs and be storable for ordinary middle-class households. For example, the value of a middle-class household's five-year income, when converted into gold bars, can be easily kept in a compact household safe. Obviously, households cannot store collateral anchors of the same value if they are commodities such as crude oil, grain or steel.
\end{itemize}

According to the above criteria, RSDM backed by precious metals or general metals such as copper can be used as a candidate for a globally acceptable moneyin the AI Era. Clearly, precious-metal-based RSDM is a symbolic weight-based money that serves as a bridge currency connecting physical assets and the AI world. RSDM can be a secure currency in the AI world that integrates intrinsic value, on-chain programmability and global neutrality. It is neither subject to the sovereign risks of fiat-collateralized stablecoins, nor as volatile as bitcoin, because precious metals are the consensus of human values and are not controlled by a single country.

\section{Discussion on the issuance forms of rsdm and the channels for the banking system to absorb and grant RSDM loans}
\label{sec:discuss}
\subsection{Potential modes of RSDM issuance}
Potential modes of RSDM issuance are listed as follows.
\begin{itemize}
    \item[(1)] Digital RSDM Hardware Wallet: built-in secure chip, enabling dual offline payment. Offline transactions can be completed via tap-to-pay without network access or a mobile phone. It displays the daily decayed balance in real time.
    \item[(2)] Paper or plastic currency (traditional physical cash): printed with a weight decay quick-reference table.
    \item[(3)] Card-style Smart Currency (Chip Cash / Programmable Cash).
    \item[(4)] Paper or plastic smart currency: paper/plastic substrate integrated with microchip and NFC module.
    \item[(5)] On-chain programmable digital currency. Similar to KAG issued by Kinesis, it is programmed to undergo weight decay at a fixed rate.
\end{itemize}

\subsection{Potential modes of banks' RSDM deposit taking and lending}
Potential modes of banks' RSDM deposit taking and lending are summarized as follows.
\begin{itemize}
    \item[(1)] Depositors deposit RSDM into banks, and the redeemed funds upon maturity remain in the form of RSDM, either physical cash or digital wallet. The interest rate and settlement currency shall be agreed upon by both parties. Interest may be settled via precious metal by weight or fiat currency, as applicable.
    \item[(2)] Taking gold-backed RSDM as an example, banks may redeem physical gold with RSDM, cover storage fees in a designated fiat currency, or place the gold in their own vaults for custodial safekeeping.
    \item[(3)] Regarding deposit balances denominated in RSDM, banks may accrue daily weight depreciation losses or purchase physical gold to offset the decayed portion for supplementary coverage.
    \item[(4)] Banks may lend out physical gold or grant loans in the form of account-based precious metals. Interest can be calculated based on precious metal weight or fiat currency.
    \item[(5)] Banks can grant loans denominated in RSDM, calculated by real-time metal weight. Such loans may be disbursed via digital wallet transfer or physical cash. Upon repayment, borrowers are required to return RSDM of equivalent weight. Interest shall be settled in fiat currency or gold, as agreed at the time of borrowing.
\end{itemize}

\section{Conclusion}
Global AI economic growth far exceeds the growth rate of traditional GDP, with the corresponding monetary demand amounting to trillions of US dollars. AI economic activities are expanding exponentially. Machine transactions in the AI ecosystem are in urgent need of a large-volume currency that supports high-frequency trading and maintains long-term value preservation. This paper focuses on activating idle private assets such as precious metals and transforming them into an excellent transaction medium for the AI era.

After analyzing transaction media including fiat-backed stablecoins, CBDCs, precious metal tokens, energy-backed currencies, and Bitcoin, we find that none of them meet the criteria for an honest money in the AI era. From the perspective of history, only precious metals are the general equivalents of the global consensus on value. At present, several precious metal tokens have tokenized physical precious metals into digital assets through blockchain technology. Each token is usually pegged at a 1:1 ratio to a specific weight of precious metal (such as 1 ounce or 1 gram), and token holders are entitled to the redemption right of the underlying precious metals. These assets integrate the value storage function of precious metals with high global liquidity. However, these tokens share a fatal defect: the lack of self-decay mechanism for metal weight. In the long term, they will inevitably evolve into credit-based currencies or face eventual collapse.

We define RSDM as the sound money for the AI ecosystem. By means of automatic weight decay, RSDM surrenders a portion of its value to cover the cumulative storage costs of collateralized precious metals over time. This automatic mechanism of partial value concession enables the residual redemption of RSDM and realizes the two-way conversion: the transformation from commodity money to symbolic currency, and from symbolic currency back to physical assets.

This paper proposes a rule framework for globally acceptable consensus currency in the AI era. It is pointed out that tokens backed by fully redeemable anchored assets represent a core feature of future sound money.The private Jiaozi, which emerged in Sichuan, China around a thousand years ago, was initially a monetized warehouse receipt. At first, it was to solve the problem of heavy and inconvenient circulation of iron money, which was a 100\% reserve and could be redeemed at the agreed business point. Because the technology at the time could not achieve the self-decay of the denomination, the promise of redemption eventually fizzled. Today, we can reimagine and restructure Jiaozi as programmable currency, and the self-depreciating RSDM mechanism is now fully technologically mature.Therefore, this paper timely proposes five issuance forms of RSDM, which fundamentally resolve the design issues of modern honest currency for both online and offline scenarios. In addition, this study further explores the mechanisms for the banking system to absorb and grant loans denominated in RSDM.

It is foreseeable that RSDM, which possesses intrinsic value and almost no logistics cost, will become the core medium for machine transactions and personal wealth storage in the AI-driven world after passing through the initial price fluctuation period.


\bibliographystyle{plainnat}
\bibliography{output}
\end{document}